\documentclass[11pt]{article}

\usepackage{graphicx}   
\usepackage{multicol}
\usepackage{color}
\usepackage{epsfig,latexsym}
\usepackage{txfonts}

\definecolor{verdon}{cmyk}{1,0.5,1,0}
\definecolor{blue}{cmyk}{0.8,0.8,0,0.}
\definecolor{red}{cmyk}{0.2,1,1,0.0}

\def\lapprox{\mathrel{\mathop  {\hbox{\lower0.5ex\hbox{$\sim$}
\kern-1.1em\lower-0.7ex\hbox{$<$}}}}}
\def\gapprox{\mathrel{\mathop  {\hbox{\lower0.5ex\hbox{$\sim$}
\kern-1.1em\lower-0.7ex\hbox{$>$}}}}}

\begin{document}

\title{\color{verdon} On the generality of the Cohen and Glashow constraints on the neutrino velocity}
\author{}

% Remove command to get current date
\date{}

\author{
F. L. Villante$^{1,2}$ and  F. Vissani$^{2}$\\
$^1${\small\em  Dipartimento di Fisica, Universit\`a dell'Aquila, L'Aquila (AQ), Italy}\\
$^2${\small\em  INFN, Laboratori Nazionali del Gran Sasso-Theory Group, Assergi (AQ), Italy}
}

\maketitle

\def\abstractname{\color{red}\bf Abstract}
\begin{abstract}
We discuss the kinematic limits 
for the process $\nu_{\mu}\rightarrow \nu_{\mu} + e^+ + e^-$ in the assumption
that neutrinos are superluminal. We derive our results by assuming that: {\em i)} it exists one reference frame in which energy and momentum are conserved; 
{\em ii)}  the Hamilton-Jacobi equation $v=dE/dp$ is valid; 
 {\em iii)} the present experimental information on the neutrino velocity at different energies are correct. 
We show that the considered process cannot be avoided unless 
{\em very} peculiar neutrino dispersion laws are assumed.
{\footnotesize }
\end{abstract}

\newpage

\paragraph{Introduction}

 The OPERA experiment recently presented a $\sim 6\, \sigma$ evidence in favor of superluminal neutrinos \cite{opera}.
By determining the time delay between $\sim 10^4$ neutrino events recorded in Gran Sasso (GS) and 
the proton collisions in CERN through a statistical comparison, they concluded that:
\begin{equation}
\delta c_{\nu} = {\rm v}_\nu-1 = \left( 2.48 \pm 0.28 \pm 0.30 \right) \times 10^{-5}
\end{equation}
The OPERA neutrino data sample covers a large energy range with 
an average energy $\langle E_\nu \rangle \simeq 17\; {\rm GeV}$. 
The experimental results do not show evidence for an energy dependence of neutrino velocity. 
By performing a two bin analysis, the experimental collaboration reported:
\begin{eqnarray}
\nonumber
(\delta c_{\nu})_1 &=& \left(2.16 \pm 0.77 \right) \times 10^{-5} \hspace{3cm}\langle E_{\nu} \rangle_1 =13.9\; {\rm GeV}\\
\nonumber
(\delta c_{\nu})_2 &=& \left(2.74 \pm 0.74 \right) \times 10^{-5}\hspace{3cm}\langle E_{\nu} \rangle_2 =42.9\; {\rm GeV}
\end{eqnarray}

If the OPERA results will be confirmed, then they will have very profound consequences on our understanding of Nature.
In the meanwhile it is important to try to falsify them, by using theoretical, phenomenological and experimental argument.
In ref.\cite{glashow} it was argued that the existence of superluminal neutrinos may be in conflict with the 
fact that high energy neutrinos propagate over the CERN-GS baseline without suffering energy losses. 
In particular, if the neutrino limiting velocity is assumed to be $(1+\delta c_{\nu})$, the process:
\begin{equation}
\nu_{\mu} \rightarrow \nu_{\mu} + e^- + e^+
\label{pair}
\end{equation} 
become possible for $E_{\nu}\ge 2\,m_{e}/\sqrt{2 \, \delta c_{\nu}} \simeq 140\, {\rm MeV}$, with the results that OPERA neutrinos should 
lose a substantial part of their energy. 
In view of the importance of this result, we would like to comment on the
generality of the underlying assumptions.
%In this note, we re-derive this result showing that this conclusion can be obtained under very general assumptions. 
Namely, we calculate the kinematic limits of the process (\ref{pair}) assuming that: {\em i)} it exists at least one reference frame in which
space and time translations and spatial rotations are exact symmetries (and, thus, energy, spatial momentum 
and angular momentum are conserved); {\em ii)} the neutrino velocity is related to its energy by 
Hamilton-Jacobi equation;  {\em iii)} the present experimental information on the neutrino velocity at different energies are correct. 

\paragraph{The basic assumptions}

 Let us assume that it exists at least one reference frame in which space and time translations and spatial rotations are exact symmetries.
This implies that energy and momentum in that reference frame are conserved, i.e. we can write for a generic physical process:
\begin{eqnarray}
\nonumber
E_{\rm ini} &=& E_{\rm fin} \\
{\bf p}_{\rm ini} &=& {\bf p}_{\rm fin}
\end{eqnarray}
where $E_{\rm ini}$ ($E_{\rm fin}$) and ${\bf p}_{\rm ini}$ (${\bf p}_{\rm fin}$) are the total initial (final) energy and momentum in the process.

 In the above assumptions, the energy cannot depend on space and time coordinates.  We can, thus, safely assume that 
the energy of a given particle ${\rm i}$ is a function of the modulus $p_{\rm i}$ of its spatial momentum, according to:  
\begin{equation}
E_{\rm i}=E_{\rm i}(p_{\rm i})
\end{equation} 
The specific form of the dispersion relation $E_{\rm i}(p_{\rm i})$ may depend on the particle type ${\rm i}$.
We assume, however, that electrons and positrons satisfy the dispersion relation provided by special relativity:
\begin{equation}
E_{\rm e} = \sqrt{p_{\rm e}^2+ m_{\rm e}^2}
\label{disSR}
\end{equation}
since there are very strong constraints on possible deviations, as it is e.g.\ reviewed in \cite{glu}.

 To infer the dispersion relation of neutrinos, we consider that the velocity of a given particle ${\rm v}_{\rm i}$ 
is related to its energy $E_{\rm i}$ by the Hamilton-Jacobi equation\footnote{As it is well known, in wave mechanics 
the Hamilton-Jacobi equation gives the group velocity of the particle wave packet.}. We have, in fact:
\begin{equation}
{\rm v}_{\rm i} = \frac{dE_{i}}{dp_{\rm i}} 
\label{HJ}
\end{equation}
This means that if we know the neutrino velocity ${\rm v}_{\nu}(p_{\nu})$ as a function of its momentum 
we can determine the function $E_{\nu}(p_{\nu})$ by performing a simple integration:
\begin{equation}
E_\nu(p_\nu)= \int_0^{p_\nu}{dq_{\nu} \; {\rm v}_{\nu}(q_{\nu})}+E_0
\label{HJInt}
\end{equation}
This information can then be used to study the kinematic limit of the process (\ref{pair}).

\paragraph{The neutrino dispersion relation}

 The observation of neutrino flavor oscillations puts very strong bounds on the possibility that neutrino dispersion relations depend on their flavor \cite{giudice}. 
We can thus assume that neutrino and anti-neutrino of different flavors have 
all the same velocities.
In this assumption, the bounds obtained from SN1987A  can be used to conclude:  
\begin{equation}
{\rm v}_{\nu}(p_{\nu}) \equiv 1 
\end{equation}
with accuracy at the level of $10^{-9}$ or more, for a 
neutrino momentum $p_{\nu} \lesssim 40$ MeV, see \cite{longo}. 
This gives automatically:
\begin{equation}
E(p_{\nu}) = p_{\nu}
\label{disLE}
\end{equation}
at low energy, where we assumed $E_0 = 0$ and neglected the neutrino mass.\footnote{
%In the following, we neglect the effect of neutrino masses and mixing. 
The  95\% CL bound on neutrino mass is 2.3 eV from Mainz \cite{mn} and 2.5 eV from Troitsk \cite{tr}
beta decay experiments respectively, and 5.7 eV from SN1987A itself \cite{sn}. The 
constant in eq.(\ref{HJInt}) is bound to be smaller than 5 eV in modulus again  
from beta decay studies \cite{vv}. All these quantities are absolutely negligible in the present context.}

The OPERA results suggest, however, that neutrino velocity deviates from 1 at high energy.
At $p_{\nu}\sim 17$ GeV, one has:
\begin{equation}
{\rm v}_{\nu} = 1 + \delta c_{\nu} 
\end{equation}
where $\delta c_{\nu} = (2.48\pm 0.28 \pm 0.30)\times 10^{-5}$.
Since there is no evidence for momentum dependence in the range $E_\nu \simeq 1 \div 40 \; {\rm GeV}$, 
we assume that $\delta c_{\nu}$ is approximately constant above a given momentum $\overline{p}_{\nu}$
which marks the transition between the low and high energy regimes.
We can, thus, integrate eq.(\ref{HJ}) obtaining:
\begin{equation}
E_\nu(p_{\nu}) = (1+\delta c_{\nu}) p_{\nu} -\delta c_{\nu} \overline{p}_{\nu}
\label{disHE}
\end{equation}
for $p_{\nu }\ge \overline{p}_{\nu}$. The transition region between the low and high energy regime 
should be approximately at $\overline{p}_{\nu}\sim 0.1 - 1\; {\rm GeV}$,  
between the energy region probed by SN1987A and that explored by OPERA.

\paragraph{The kinematic of the process}

 By using energy-momentum conservation and the dispersion relations (\ref{disSR}), (\ref{disLE}) and (\ref{disHE}),
we can show that the process (\ref{pair}) is allowed by kinematic constraints.
We assume that the particles in the initial and final state are collinear. 
We also assume that the electron and the positron have the same momentum $p_{\rm e}$.
In this assumption, we obtain:
\begin{eqnarray}
\nonumber
E_{\nu} - E'_{\nu} &=& 2 \sqrt{m_{\rm e}^2+p^2_{\rm e}} \\
p_{\nu} - p'_{\nu} &=& 2 p_{\rm e} 
\end{eqnarray}
We now consider the case in which the initial neutrino momentum is larger than $\overline{p}_{\nu}$
while the final neutrino momentum is negligible $p'_{\nu}\simeq 0$. We thus have:
\begin{eqnarray}
 (1+\delta c_{\nu}) p_{\nu} -\delta c_{\nu} \overline{p}_{\nu}   &=& 2 \sqrt{m_{\rm e}^2+p^2_{\rm e}} \\
p_{\nu}  &=& 2 p_{\rm e} 
\end{eqnarray}
By solving the above equations, we obtain a condition:
\begin{equation}
p_{\nu} \ge p_{\rm thr.} = 
%\frac{\delta c_{\nu}\,\overline{p}_{\nu}  + \sqrt{ \left( \delta c_{\nu}\,\overline{p}_{\nu} \right)^2+ 8 \, m_{\rm e}^2 \, \delta c_{\nu}}}
%{2\;\delta c_{\nu}}
\frac{\overline{p}_{\nu} +\sqrt{\,\overline{p}_{\nu}^2 +8\, m_{\rm e}^2\,/\delta c_{\nu} }  }{2}
\label{final}
\end{equation}
where we neglected terms proportional to $\delta c_{\nu}^2$.
By using $\overline{p}_\nu = 100\; {\rm MeV}$ and $\delta c_{\nu} = 2.5 \times 10^{-5}$, we obtain $p_{\rm thr.} \simeq 200 \; {\rm MeV}$.
Assuming transition at larger energies, e.g. choosing $\overline{p}_\nu\ge 1\;{\rm GeV}$,
the above equation gives $p_{\rm thr.}\sim \overline{p}_\nu$. All this shows that neutrinos produced in OPERA 
should be  affected by pair production energy losses.

The above conclusion has been obtained by using  eqs.~(\ref{disLE}) and (\ref{disHE}) which corresponds to assuming a 
relatively fast transition between the low and high energy regime, as it is done e.g. in \cite{caccia}.
It continues to hold with small changes in all scenarios in which velocity is a monotonic growing function
of momentum that has $\delta c_{\nu} (p_{\mbox{\tiny OPERA}})\sim 2.5\times 10^{-5}$, 
like those developed before OPERA results \cite{previous}.

In order to avoid pair production, we need that:
\begin{equation}
E_{\nu}(p_{\nu})^2- p_{\nu}^2 \le (2 m_{\rm e})^2
\end{equation}
 at OPERA energies. This can be obtained only if we assume that neutrinos moves slower than light in the intermediate region 
$p_\nu\sim 0.1-1 \mbox{ GeV}$, so that the contribution from a negative $\delta c_{\nu}$ in integral (\ref{HJInt}) compensates the 
effect of the $\delta c_{\nu} > 0$ reported by OPERA, for instance
\begin{equation}
\delta c_\nu=
\left\{
\begin{array}{cl}
-2.728\times 10^{-3} & \mbox{ when } p_\nu=0.1-1 \mbox{ GeV}\\
+2.480\times 10^{-5} & \mbox{ when } p_\nu=1-100\mbox{ GeV}\\
0 & \mbox{ otherwise}
\end{array}
\right.
\end{equation}
where we set $\delta c_{\nu}$ to zero below 100 MeV in order not to contradict SN1987A findings
 but also  above 100 GeV in order to account for the observation of high energy muon neutrinos \cite{glashow}. 
The constant are arranged so that $E_\nu< p_\nu$ for $p_\nu  = 0.1 - 100$~GeV,  and $E_\nu=p_\nu$ elsewhere. 
 Such a possibility seems implausible, but can be possibly further tested studying neutrino velocity in a different range of momenta and in principle considering processes, as 
the emission of neutrino pairs $e^-\to e^- \nu_\mu \bar\nu_\mu$ in vacuum.
In short, it seems that the very special possibilities where the constraints of Cohen and Glashow can be evaded will be directly constrained 
by present and future experimental data (e.g. by K2K, MINOS, and T2K that have experimental data in the intermediate energy region).

\paragraph{Conclusion}

 We have shown that the conclusion of \cite{glashow} who argued that 
the existence of superluminal neutrinos may be in conflict with the fact that high energy neutrinos 
propagate from CERN to Gran Sasso without suffering severe energy losses 
follows from very general assumptions. 

Namely, the kinematic limit for the process (\ref{pair}) can be derived by only assuming that:
{\em i)} it exist one reference frame in which
space and time translations and spatial rotations are exact symmetries (and, thus, energy, spatial momentum 
and angular momentum are conserved); {\em ii)} the neutrino velocity is related to its energy by 
Hamilton-Jacobi equation;  {\em iii)} the present experimental information on the neutrino velocity
at different energies are correct. 

By considering the SN1987A bounds on neutrino velocity at low energy and assuming that the neutrino 
velocity above a certain momentum ${\overline p}_{\nu} \sim 100 \; {\rm MeV}$ 
(up to the OPERA neutrino energy range) is approximately constant, we derived 
the condition (\ref{final}) that shows that neutrinos produced in OPERA should be affected
 by pair production energy losses.
In conclusion, it appears difficult to evade the argument of \cite{glashow}
unless energy-momentum conservation is broken in any reference frame and/or 
very peculiar neutrino dispersion laws are considered. 

\section*{\sf  References}
\def\refname{

\end{document}
\vskip-1cm}
\baselineskip=1.15em


\begin{thebibliography}{99}

\bibitem{opera}
T.~Adam {\it et al.} [OPERA Collaboration],
``Measurement of the neutrino velocity with the OPERA detector in the CNGS beam,''  
  arXiv:1109.4897 [hep-ex] V1.

\bibitem{glashow}
  A.~G.~Cohen, S.~L.~Glashow,
  ``New Constraints on Neutrino Velocities,''
  arXiv:1109.6562 [hep-ph].

\bibitem{glu}
S.~R.~Coleman, S.~L.~Glashow,
  %``High-energy tests of Lorentz invariance,''
  Phys.\ Rev.\  {\bf D59 } (1999)  116008.
  
  
\bibitem{giudice}
G.~F.~Giudice, S.~Sibiryakov, A.~Strumia,
  ``Interpreting OPERA results on superluminal neutrino,''
  arXiv:1109.5682 [hep-ph].


\bibitem{longo}
The analysis of M.~J.~Longo,
 %``TESTS OF RELATIVITY FROM SN1987a,''
Phys.\ Rev.\  {\bf D36 } (1987)  3276
obtains $|\delta c_\nu|<2\times 10^{-9}$.
The bound becomes more stringent by a factor of
$\sim 6$ by modeling the propagation of the shock wave 
as in W.~D.~Arnett, Astroph. J. {\bf 331} (1988) 377 and 
and including the observation of A.~Jones of the IAU Circ. 4316 (1987).


  
\bibitem{mn}
Ch. Kraus {\em et al.}, Eur. Phys. J. {\bf C40} (2005) 447.

\bibitem{tr}
V.~M. Lobashev {\em et al.}, Phys.Lett. {\bf B460} (1999) 227.

\bibitem{sn}
T.~J.~Loredo, D.~Q.~Lamb,
  Phys.\ Rev.\  {\bf D65 } (2002)  063002;
G.~Pagliaroli, F.~Rossi-Torres, F.~Vissani,
  Astropart.\ Phys.\  {\bf 33 } (2010)  287.



\bibitem{vv}
F.~Vissani,
  %``The Beta spectrum in presence of background potentials for neutrinos,''
  Phys.\ Lett.\  {\bf B413} (1997) 101, see Eq.~(26).


\bibitem{caccia}
G.~Cacciapaglia, A.~Deandrea, L.~Panizzi,
  ``Superluminal neutrinos in long baseline experiments and SN1987a,''
  arXiv:1109.4980 [hep-ph].



\bibitem{previous}
J.~R.~Ellis, N.~Harries, A.~Meregaglia, A.~Rubbia, A.~Sakharov,
  %``Probes of Lorentz Violation in Neutrino Propagation,''
  Phys.\ Rev.\  {\bf D78 } (2008)  033013.

\end{thebibliography}
